\documentclass[12pt]{article}
\usepackage[tbtags]{amsmath}
\usepackage{graphicx}
\usepackage{amssymb}
\usepackage{epstopdf}
\usepackage{named}
\usepackage{jme}
\renewcommand{\baselinestretch}{1.5}

\providecommand{\e}[1]{\ensuremath{\times 10^{#1}}}

\begin{document}
\title{Monomer Abundance Distribution Patterns as a Universal Biosignature: Examples from Terrestrial and Digital Life}
\author{Evan D. Dorn$^{1}$, Kenneth H. Nealson$^{2}$, Christoph Adami$^{1,3}$}
\date{}
\maketitle
\begin{center}
{\it$^{1}$Digital Life Laboratory 136-93, California Institute of Technology\\
$^{2}$Department of Earth Sciences, University of Southern California\\
$^{3}$Keck Graduate Institute of Applied Life Sciences}
\end{center}

Corresponding Author: \\
 Christoph Adami \\
Keck Graduate Instititute\\
 Claremont, CA 91711\\
Email: adami@kgi.edu\\
 Phone: (+)909-607-9853\

\pagebreak

\begin{abstract} 
 Organisms leave a distinctive chemical signature in their environment
 because they synthesize those molecules that maximize their fitness. 
 As a result, the relative concentrations of related chemical monomers
 in life-bearing environmental samples reflect, in part, those
 compounds' \textit{adaptive utility}.  In contrast, rates of molecular synthesis
 in a lifeless environment are dictated by reaction kinetics and thermodynamics,
 so concentrations of related monomers in abiotic samples tend
 to exhibit specific patterns dominated by small, easily formed, low-formation-energy
 molecules.  We contend that this distinction can serve as a universal
 biosignature: the measurement of chemical concentration ratios that
 belie formation kinetics or equilibrium thermodynamics indicates the likely
 presence of life.  We explore the features of this biosignature as
 observed in amino acids and carboxylic acids, using published data from
 numerous studies of terrestrial sediments, abiotic (spark, UV, and 
high-energy proton) synthesis experments, and meteorite bodies. 
We then compare these data to the results of
 experimental studies of an evolving \textit{digital life} system.  We observe the
 robust and repeatable evolution of an analogous biosignature in a
 digital lifeform, suggesting that evolutionary selection necessarily
 constrains organism composition and that the monomer abundance
 biosignature phenomenon is universal to evolved biosystems.\end{abstract}
 
\section{Introduction}

The environmental concentrations of related chemical species carry
information about their origin and synthesis.  Relative rates of
synthesis for individual chemical species differ between biotic and
abiotic sources.  Where molecules are synthesized by abiotic
processes, rates of formation are constrained by the laws of
thermodymamics and kinetics, resulting in a distribution of molecules
dominated by low molecular weight and kinetically-allowable species.  Organisms,
on the other hand, contain catalysts (e.g., in terrestrial biota,
enzymes) and expend energy to synthesize specifically those molecules
they need for survival and competition.  In the presence of life,
therefore, some specific complex and high-formation-energy molecules are
synthesized rapidly because they convey a fitness benefit.  If the population
is large enough, the organisms' chemical composition can become 
measurable in, or even dominant over, the chemical composition of the
environment.  In effect, the action of evolution on genomes can imprint a 
signature on environmental chemistry because fit genomes synthesize 
molecules at rates that reflect their utility for propagating the genome within the given environment, 
thus altering the molecules' distribution in the environment. 
If the population is large enough to contribute to bulk concentration and if 
these compounds are not quickly destroyed (non-biologically 
mediated diagenesis), biosynthesis (i.e., biogenesis) may be a 
significant factor governing the observed concentrations in a sample.
Similarly, biologically-mediated diagenesis may affect environmental
concentrations in a way that reflects evolved metabolic activity and,
therefore, selection.

The distribution of a set of molecules in an environmental sample,
therefore, may indicate the presence or absence of life.  The principle that 
patterns of molecular abundance within a family of molecules 
to distinguish abiotic and biotic sources of those molecules has been
discussed previously~\cite{Lovelock:1965,McKay:2002,McKay:2004,SummonsRE_etal:2008,ShapiroR_Schulze-MakuchD:2009,DaviesPCW_etal:2009}.  More specifically, the principle that
biological systems employ a discontinuous subset of the possible molecules
in a family of biochemicals while abiotic syntheses produce a continuous 
spectrum of molecules, has been observed in many of these discussions and 
has been termed the ``Lego Principle" by McKay \cite{McKay:2004,DaviesPCW_etal:2009}. 
While indeed life uses only a subset of the possible molecules in many chemical families, 
a discontinous subset as a feature of life is not key to the argument we outline here.  
In at least some cases, as we will discuss later, the full spectrum of molecules are observed 
in both abiotic and biotic contexts.  Thus it is only important to our
argument that the abiotic and biotic abundance patterns be distinguishable, which
we believe to be likely even in the case where all molecules in the family
are observed in a sample.
 
We refer to the environmental distribution of molecular abundances as the ``monomer abundance distribution biosignature'' (MADB) 
and hypothesize that it is universal to all forms
of life and collections of monomers they employ.   While in principle we
may observe such a biosignature in any set of chemical measurements, in
practice it is easiest to demonstrate by comparing the relative
concentrations of chemicals within a single family of related molecules
(such as amino acids) where comparison of synthesis rates is arguably
valid.  If our hypothesis is true, it carries implications both for the
understanding of evolving biochemistries and the detection of
extraterrestrial life, where the biochemistry of a putative ecosphere is
not \textit{a priori} known.

Of course, if significant periods of time elapse between the formation 
of the molecular distribution via biotic processes and its analysis, 
diagenesis can obscure any biosignature because different monomers may 
degrade at different rates. For example, the low molecular weight amino 
acids can come to predominate in natural sediments over 
time \cite{AbelsonPH_HarePE:1969,ElsterH_etal:1991} simply because 
they are more stable, so that  the original distribution 
is highly altered. The same is true of the signatures in meteorites, 
which must reflect the high-energy processing the organic compounds 
have experienced over the last 4.5 billion years.

While we apply the MADB to the challenge of detecting life
in an extraterrestrial environmental sample, the use of monomer
distributions as a diagnostic tool for other purposes is
well-established.  Fatty acid and phospholipid distributions have long
been employed in the study of microbial diversity and other fields of
biogeochemistry; see e.g. \cite{WhiteDC_etal:1997,ZellesL_BaiQY:1994,SundhI_etal:1997},
and amino acid profiles are frequently used as indicators of the diagenetic state of organic matter
\cite{HedgesJI_OadesJM:1997,DauweB_MiddelburgJJ:1998,ColomboJC_etal:1998}.  
Amino acid profiles have also been used to distinguish between the synthesis pathways underlying
abiotic amino acid formation in carbonaceous chondrites \cite{BottaO_etal:2002,LernerNR_etal:1993}.   
The relative frequencies of individual nucleotides in gene sequences are
highly variable, but have statistical properties that can
distinguish kingdoms and smaller families 
\cite{SchultesEA_etal:1997,GorbanAN_etal:2003}. Also,
patterns of nucleic acids have been shown to reflect selective 
constraints such as RNA folding, thereby
distinguishing functional from non-functional RNA 
\cite{SchultesEA_etal:1999}, as well as the likely availability of 
particular amino acids in
the prebiotic environment \cite{BrooksDJ_etal:2002}.   Geochemical
profiles have even been employed previously to detemine whether
hydrocarbon compounds in Earth's mantle are of a biotic or abiotic
source, through comparison to the contents of meteorites and known
abiotic syntheses \cite{SugisakiR_MimuraK:1994}.

Amino and carboxylic acids are familiar chemical families in which 
terrestrial biochemistry exhibits a clear signature of life's
presence, and we collect
and compare here data from a number of literature sources to demonstrate
the effect. The statistical significance of the monomer abundance
biosignature in amino acids has been previously studied \cite{DornED_etal:2003}, 
and we review here the most salient results.   In this study, we
conduct a more thorough analysis of amino acids and include saturated
monocarboxylic acids  in order to demonstrate that the
MADB phenomenon as observed in the terrestrial
biosphere is not limited to only one monomer family.

Any proposed biosignature should be testable not only against terrestrial,
but also extraterrestrial life.  To quote Maynard Smith (1992)\nocite{SmithJM:1992}: ``So far,
we have been able to study only one evolving system and we cannot wait
for interstellar flight to provide us with a second. If we want to
discover generalizations about evolving systems, we will have to look at
artificial ones.'' We would like to go a little further to suggest that,
in order to \textit{detect} extraterrestrial life, we should have an
understanding of the fundamental dynamics of living and evolving
systems, independent of the organisms' particular substrate (i.e.,
chemistry). \textit{Digital life} \cite{AdamiC:1998,RayTS:1992} is an artificial form of
life that provides just such an instance, and it has been used
successfully as an experimental platform for the study of evolutionary
dynamics \cite{AdamiC:1998,AdamiC_etal:2000,BellG:2001,ChowSS_etal:2004,LenskiRE_etal:1999,LenskiRE_etal:2003,OfriaC_etal:2002,OfriaC_etal:2003,WilkeCO_AdamiC:2002,WilkeCO_etal:2001,YedidG_BellG:2001}.  In a digital life
system, self-replicating computer programs evolve within a digital
ecosystem, enabling precise and controlled studies of evolutionary
dynamics.

\subsection{The Avida Environment}

The Avida Digital Life Platform \cite{OfriaC_WilkeCO:2004}
is ideal for our purposes, since it is a controlled and understood system unrelated to
terrestrial life.  A digital system also makes possible repeated
experimentation with evolution.  An introduction to digital life can be
found in Adami (1998) and Wilke and Adami (2002)\nocite{AdamiC:1998}\nocite{WilkeCO_AdamiC:2002}, while Adami (2006) \nocite{Adami2006} should be consulted for a more recent overview of research conducted with this form of life.

In the digital world of Avida, computer programs of varying length
encoded in a simple language replicate themselves and compete for
processor time and physical space. Avidians evolve by mutation and
selection of a self-replicating genome with access to a ``chemistry'' of
simple different instructions (the monomers; 29 were available in the
variant used for this study). Typically, a population is seeded with a
primitive organism made from only 20 of those instructions.  As the
organism replicates, mutations (including copy errors, spontaneous point
mutations, and insert and divide mutations) are imposed randomly on all
organisms in the environment at rates configurable by the experimenter. 
Mutations and insertions create a new instruction (randomly chosen from
the library of 29 with equal probability for each) at the target
location.

In order to replicate, an avidian needs energy in the form of CPU
(central processing unit) time.  Each organism can increase its fraction
of CPU time by performing calculations on a stream of random input
numbers. These calculations (computational tasks) are the analogue of
metabolic reactions (in biochemistry), and the code evolved to perform
these tasks is analogous to the genetic code for metabolic enzymes.  The
simple self-replicator used as a progenitor (ancestor) organism does not
include code to complete any mathematical tasks.  As genotypes evolve
computational genes, they receive a higher proportion of CPU time and
quickly come to dominate the population.

We examined the relative distributions of computer instructions (the
fundamental ``monomers" in Avida) of numerous evolved and evolving
populations to characterize the nature and robustness of the abundance
patterns they form in response to evolutionary pressures. Avida
organisms (``avidians'') consist of a single genome: a string of computer
instructions.  These genomes evolve as a result of externally-imposed
mutations and selection pressures (the environment). The computer
instructions that compose avidians are analogous to the amino
acids, fatty acids, or other compounds that compose familiar biota in
the sense that they are selectable, meaning only those instructions that
carry a fitness advantange will be reproduced in future generations. 
Therefore, any general fitness advantage or adaptive utility of a single
instruction should be visible in its bulk concentration in an evolved
population.

We believe that Avida represents a sufficiently analogous model of biological 
evolution to test the MADB hypothesis.  In both a biosphere and in the Avida simulator,
self-replicating lifeforms are composed of one or more sets of monomers (e.g. amino acids, 
carboxylic acids, computer instructions).
In the absence of life, abiotic formation processes
will entirely determine the relative ratios of these monomers' concentrations, and we
possess models and measurements of abiotic processes in both Avida and in organic chemistry. 
In the presence of life, however, the composition of the organisms will affect
the concentration ratios of those monomers in the environment.   In both Avida and the 
biosphere, competition will result in selection of the organisms, and this selection will act
to optimize the composition of the organisms.  For example, selection is unlikely to produce
organisms whose proteins all comprise 95\% glycine, since such proteins would not likely
be functional, and certainly not optimally adapted.  Likewise, in Avida, an adapted organism's
instruction sequence is unlikely to contain 95\% NOP-A instructions, since such a sequence
would be inefficient at best.  As a result, we can test in both the biosphere and in Avida the
hypothesis that the concentration profile observed in an adapted organism - and by implication its environment - is unlikely to be the same we would observe in the absence of life.

In the biosphere, a high abundance of heavy amino acids like glutamate (relative to glycine) is
diagnostic for life, because only a selected metabolism can account for such a 
thermodynamically improbable ratio.   Likewise, in an Avida array, the presence of 
too-high or too-low concentrations of specific instructions can indicate the presence of 
selective pressure on the digital analogue of metabolism, since the Avida simulator inserts all
instructions in equal proportion.
 
\section{Materials and Methods}

\subsection{Terrestrial Biochemistry}

\subsubsection{Data Sources}

We compiled a database of measurements of related monomer concentrations
(131 samples of amino acids, 31 samples of straight-chain monocarboxylic
acids) from a variety of publications including studies of formation
synthesis, terrestrial water columns and sediments, and meteorites.

Our set of abiotic amino acid data included thirty measurements in
total, fifteen from tholin and spark-synthesis experiments 
\cite{KhareBN_etal:1986,McDonaldGD_etal:1994,McDonaldGD_etal:1991,MillerSL_UreyHC:1959,RingD_etal:1972,SchlesingerG_MillerSL:1986}
one from a report of UV photolysis in deep-space conditions \cite{Munoz-CaroGM_etal:2002}, one of
amino acid synthesis in a proton-irradiatiated mixture \cite{TakanoY_etal:2004},
seven from the Murchison meteorite
\cite{CroninJR_etal:1981,CroninJR_MooreCB:1976,CroninJR_etal:1980,CroninJR_etal:1979,EngelMH_MackoSA:1997,EngelMH_etal:1990,EngelMH_NagyB:1982,KvenvoldenKA_etal:1970}
and six from two other carbonaceous chondrites \cite{CroninJR_etal:1979,ShimoyamaA_etal:1979}. 
Based on composition and racemic mixture, the
amino acids from these meteorites are generally agreed to be of abiotic
origin \cite{CroninJR_etal:1994,EngelMH_MackoSA:1997}.

The complementary dataset of amino acids from the terrestrial biosphere
included 125 measurements of extracts from a variety of terrestrial
sediments, soils, and water columns
\cite{ColomboJC_etal:1998,CowieGL_etal:1992,DauweB_MiddelburgJJ:1998,HedgesJI_etal:2000,HorsfallIM_WolffGA:1997,KeilRG_etal:1998,KeilRG_FogelML:2001,Kielland_K:1995}.  These measurements include samples at a variety of ages and diagenetic condition ranging from fresh soil \cite{Kielland_K:1995} and river waters \cite{HedgesJI_etal:2000} through ocean sediments up to 95 - 170 years old \cite{CowieGL_etal:1992}.   

Our database of carboxylic acids from abiotic sources includes two measurements from
spark synthesis studies \cite{ShimoyamaA_etal:1994,YuenGU_etal:1981} and
three from meteorite sources \cite{ShimoyamaA_etal:1986,NaraokaH_etal:1996,LawlessJG_YuenGU:1979}.  
All five sources report only the
concentrations of short-chain carboxylic acids (C12:0 and shorter).

We rejected a few otherwise interesting measurements of amino acids
including simulated Martian conditions \cite{AbelsonPH:1965}, early
spark-synthesis studies \cite{MillerSL:1953}, Fischer-Tropsch type (FTT) and ``thermal" syntheses
\cite{AndersE_etal:1973,HayatsuR_etal:1971} as well as amber-encased insects
\cite{WangXS_etal:1995} because they reported quantified measurements of too few 
of the amino acids (Gly, Ala, $\beta$-Ala, etc.) for useful comparison to other studies.

We compiled two different datasets of carboxylic acid measurements from
terrestrial sources.   We include only the straight-chain saturated
monocarboxylic acids so that the data can  be compared to sources of
data on abiotic formation that report only this subfamily of compounds. 
The first dataset includes twelve measurements of the same low-weight acids as
the abiotic dataset, all taken from coastal marine sediments \cite{ShimoyamaA_etal:1991}.  
A second group contains twenty-four measurements of
longer chain carboxylic acids C14:0 through C28:0 in a variety of
sediments and soils \cite{ZellesL_BaiQY:1994,WakehamSG:1999,ColomboJC_etal:1996,BaathE_etal:1992}.

\subsubsection{Data Treatment}

To combine the disparate sources of data into unified sets for
comparison, we applied the following protocol:

\begin{enumerate}
\item All data were converted to molar units.
\item Residues reported as ``coeluted'' (e.g., glycine and glutamate
reported together as GLX) were both given the reported value.
\item ``Trace'' values were given a concentration of 0.
\item Where enantiomers were reported separately, their sum was used in
a combined column.
\item Values reported as ``approximate'' or ``estimated'' were used
unchanged.
\item Where values were reported as ``$<$ X'', we used X.
\end{enumerate}

It should be noted that while while enantiomeric excess has long been
recognized as an important biosignature, we ignore it here because it was not 
measured in the vast majority of our data sources and is not the focus of
this work. 

In addition, we applied the following steps to the amino acid
and long-chain carboxylic acid datasets:
\begin{enumerate}
\setcounter{enumi}{6}
\item Where a clear statement could be found in the text that a search
was exhaustive (e.g., ``all other residues were absent or present only at
trace levels''), unreported measurements were assigned a zero.
\item Other unreported values were left blank and not included in
averages.
\item Columns with more than 50\% blank values in any one category were
eliminated from the analysis.  (e.g., Sarcosine is quantified in many
spark-synthesis and meteorite studies, but never in sediment studies.)
\end{enumerate}

The reduced amino acid database included eleven residues (from an
initial thirty) that were quantified with sufficient frequency in both
the biotic and abiotic categories: Gly, Ala, $\beta$-Ala, Aba, Ser, Ile, Leu,
Asp, Glu, Thr, Val.  The reduced long-chain carboxylic acid dataset
included all straight-chain acids from C9:0 through C28:0, except C27:0.
 The abiotic component of the short-chain carboxylic acid dataset did
not contain sufficient overlapping data for reasonable presentation of
averages, so we have presented each measurement individually.

Each dataset was then normalized to a value appropriate for that
category.  All amino acid measurements were normalized to the mole
concentration of glycine in that sample.  The short- and long-chain
carboxylic acid datasets were normalized to the concentrations of
propionic acid (C3:0) and palmitic acid (C16:0), respectively.

\subsection{Digital Life}

\subsubsection{Evolved Monomer Abundances in Terminal Populations}

Our first experiment examined the final distribution of instructions in
350 evolved populations that varied in physical parameters and initital
conditions.  In each run, we populated a grid of 3600 cells with one of
two hand-coded ancestor genotypes, and allowed the population to evolve
for 4000 generations.  We quantified the bulk frequency of each of the
instructions in the population every ten generations.

We used two different ancestor genotypes, (carefully written to differ
as much as possible from each other), to control for the effect of
initial conditions on the final evolved biosignature.  The two
progenitors were Avida's default 20-instruction ancestor, which uses the
instructions SEARCH-F, JUMP-B and INC for the primary functions of flow
control and replication, and an alternate 55-instruction ancestor that
replaces those instructions with CALL, RETURN, SEARCH-B, SHIFT-L, DEC,
and JUMP-F.  Both organisms are shown in Table 1.

We also varied the copy-mutation rate in seven levels ranging by factors
of two from 0.00125 to 0.08 mutations per copied instruction.  At each
combination of ancestor and mutation rate, we performed twenty-five runs
that differed only in random seed.

\subsubsection{Time-Evolution of a Biosignature}

We conducted a second experiment in Avida to show the time-course of the
evolving MADB as a formerly abiotic
environment becomes dominated by incident life.   We generated an
abiotic population by initializing the same 3600-cell Avida grid with
randomly-generated, nonviable (meaning nonreplicating) genomes.  We
created a lethal environment by bombarding the entire population with a
high rate of point mutations. (A lethal mutation rate is so high that
information cannot be maintained in a self-replicating genome;
eventually the gene for self-replication is lost.)  For the duration of
the experiment, we periodically seeded this population with viable
(mutation-free) progenitor organisms. This is analogous to a steady
influx of space-borne spores onto a planet with a lethal environment and
no existing biology. Then, we progressively stepped down the rate of
point mutations until it reached a level where the organisms could
survive and a population began to replicate and evolve. The population
was allowed to evolve until the distribution of instructions had
stabilized for several hundred generations. Then the point mutation rate
was stepped back up until the population died and the randomizing effect
of the mutations returned the instruction distribution to its
``prebiotic'' state.

\subsubsection{Data Treatment}

Of the twenty-nine instructions in Avida's library, we exclude one
instruction, NOP-A, from the analysis.  NOP-A is used by the system to
initialize empty memory in dividing organisms and is therefore greatly
over-represented in comparison to the other instructions, and fluctuates
significantly as organisms execute the ALLOCATE instruction.
Furthermore, NOP-A's base rate of synthesis and availability is strongly
genotype-dependent.  It cannot be estimated a priori and we cannot make
reasonable statements as to whether its abundance reflects biotic or
abiotic processes, so we therefore compare only the remaining 28
instructions.

We quantified terminal abundances in evolved populations by averaging
the last ten measurements, to provide data smoothing.  Avidians in the
process of replicating, and occasional nonviable mutations, cause small
instantaneous fluctuations in population abundance of the individual
instructions.

\section{Results}

\subsection{Terrestrial Biochemistry}

\subsubsection{Amino Acids}

Glycine and alanine, which have low molecular weight and whose synthesis
pathways are kinetically favorable, dominate mixtures of amino
acids synthesized by abiotic processes. Some small non-protein amino
acids, including sarcosine and beta-alanine, have been observed in
synthesis experiments \cite{KhareBN_etal:1986,McDonaldGD_etal:1994,McDonaldGD_etal:1991,MillerSL:1955,MillerSL_UreyHC:1959,Munoz-CaroGM_etal:2002,SchlesingerG_MillerSL:1986}. Other amino acids are
present only in trace concentrations, if at all 
\cite{McDonaldGD_etal:1994,McDonaldGD_etal:1991,MillerSL:1955,Munoz-CaroGM_etal:2002}. This pattern
reflects the thermodynamics and kinetics of amino acid synthesis and is
remarkably consistent regardless of the specific nature of the synthesis
environment.  For example, the alpha amino acids are likely formed via
the Strecker (cyanohydrin) synthesis \cite{MillerSL:1955},
which presumes the existence of precursor molecules that include the
sidechain. The formation of heavier amino acids would require the prior
formation of the larger sidechains: since these reactions would include
their own kinetic barriers, the availability of the larger precursors
and ultimately the final yields of the larger amino acids are expected
to be lower.   In addition, larger sidechains have a greater number of isomers. 
Therefore, unlike biotic synthesis where an enzyme catalyzes formation of a specific
desired isomer, we would expect an abiotic synthesis to produce smaller
quantities of many isomers.

Some of the heavier protein amino acids such as Tyr and His have been detected 
in FTT and high-temperature syntheses \cite{AndersE_etal:1973,HayatsuR_etal:1971}.
However, we are unaware of any measurements of these molecules in abiotic
synthesis in quantities that approach the concentration of Gly.  Phe and Thr are 
not seen in abiotic synthesis.

Amino acids in terrestrial samples show a more varied distribution
dominated by the protein amino acids in roughly equal proportions.
Figure 1 shows the relative abundances of a set of twelve amino acids
measured in a variety of biotic and abiotic samples.  The amino acids
are plotted in ascending order of $\Delta$G$_{r}$ (Gibbs free energy of synthesis)
at 18$^{\circ}$C as reported by Amend and Shock (1998)\nocite{AmendJP_ShockEL:1998} to show both the trend
among abiotic synthesis products towards smaller and ``simpler'' compounds
and the biosynthesis of complex, expensive compounds. The
``Sediment'' curve represents terrestrial sediment and water column
extractions. ``Meteorite'' represents extractions from three carbonaceous
chondrites, and ``Synthesis'' is an average of fifteen spark-synthesis
experiments in a variety of atmospheres and a single extraction from
amino acids synthesized via UV photolysis on analogues of interstellar
ice. The Synthesis curve serves as an abiotic baseline: the pattern of
amino acid concentrations we expect when life is not present. Except for
a single anomaly (high glutamate concentration), the meteorite curve
exhibits the same pattern, while the biologically-generated Sediment
curve differs significantly, exhibiting high concentrations of a number
of the larger amino acids. The measured concentrations of the
heavier protein amino acids vary greatly in biotic samples, but are
consistently higher than is ever observed in abiotic sources.

\begin{figure}
\centering
\includegraphics[width=10cm]{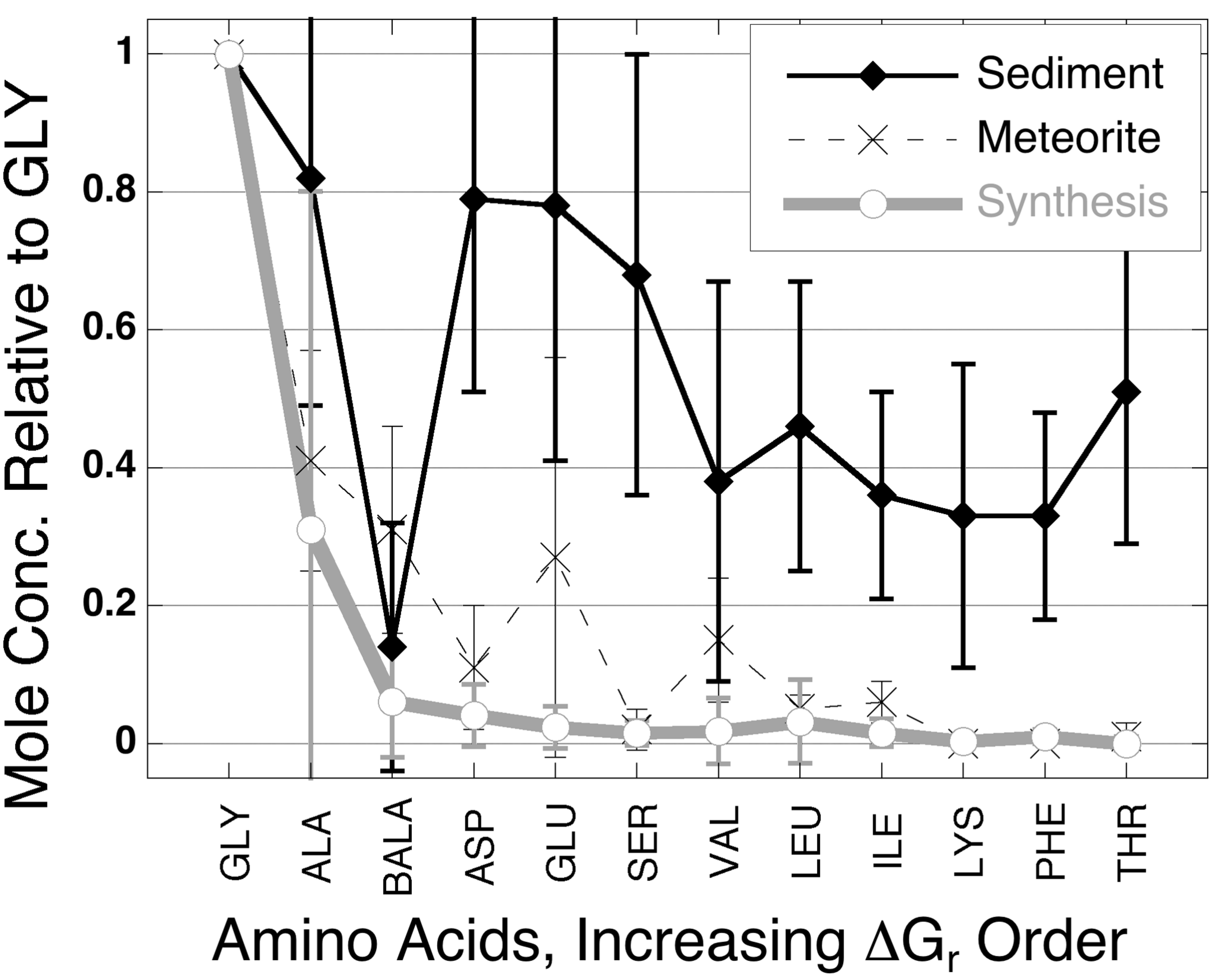}
\caption{Average patterns of amino acid abundances relative to glycine,
compared between biotic (Sediment $n$=125) and abiotic (Meteorite $n=15$,
Synthesis $n=16$) sources. Error bars are one standard deviation.}
\label{Figure 1}
\end{figure}

This analysis only shows a small selection out of hundreds of possible
amino acids.  A quantification of all possible low-molecular-weight
amino acids would show several, such as sarcosine, that are non-zero in
the abiotic baseline but low or zero in the biotic curves, similar to
beta-alanine in Figure 1. Many heavier amino acids would not be present in
either case.  Therefore, the complete amino acid biosignature for
terrestrial organisms would appear graphically as a few dozen peaks for the
protein amino acids, other biologically important amino acids such as
synthesis intermediates and neurotransmitters, and major breakdown products
in a large spectrum of hundreds of possible, but absent, amino acids.   This
sparse subset and its utility as a biosignature is referred to by McKay 
as the Lego Principle \cite{McKay:2004}.   Fig. 1 is a subset of the hypothetical
 biosignature plot that includes all possible amino acids found in either
 biotic or abiotic contexts.

\subsubsection{Carboxylic Acids}

Preparations from spark-synthesis studies exhibit carboxylic acids of
low carbon number: only acids C6:0 and smaller are seen in significant
quantities, and acids larger than C12:0 are not found at all in most
experiments \cite{ShimoyamaA_etal:1994,YuenGU_etal:1981}. Figure 2 shows the relative
concentrations of the first few acids measured in two such experiments
and as measured in three meteorites, compared to an average of twelve
measurements of terrestrial sediments. We observe in the Synthesis data
a consistent exponential decrease in concentration with increasing
molecular weight, while the Sediment curve shows a much more even
distribution.   The meteorite data fall between the two extremes, with a
shallower exponential decrease and some variability in the relative
concentration of acetic acid, probably due to its relatively high
volatility and possible outgassing in space or during atmospheric entry.
 We contend that the pattern reflects the same fundamental processes
that are at work in the amino acid data. As molecules become larger,
formation energies are higher, synthesis kinetics generally become less
favorable, and the number of possible conformations is greater, reducing
the expected relative concentration of any particular conformer. The
data from the meteorites implies a probable abiotic origin with a small
amount of contamination by terrestrial material.  We are not aware of
any significant quantities of carboxylic acids larger than C12:0
recovered from non-contaminated meteorites.  

Long-chain (up to C33) 
lipid compounds have been created in Fischer-Tropsch type (FTT) syntheses 
at high (200 $^\circ$C and greater) temperatures, but such experiments report no bias 
toward any particular carbon number \cite{McCollomTM_etal:1999,RushdiAI_SimoneitBRT:2001}.  
Therefore the bias towards even-numbered carbon chains in carboxylic acids
serves as an MADB even when long-chain-producing FTT processes are 
included in the abiotic baseline.

\begin{figure}
\centering
\includegraphics[width=12cm]{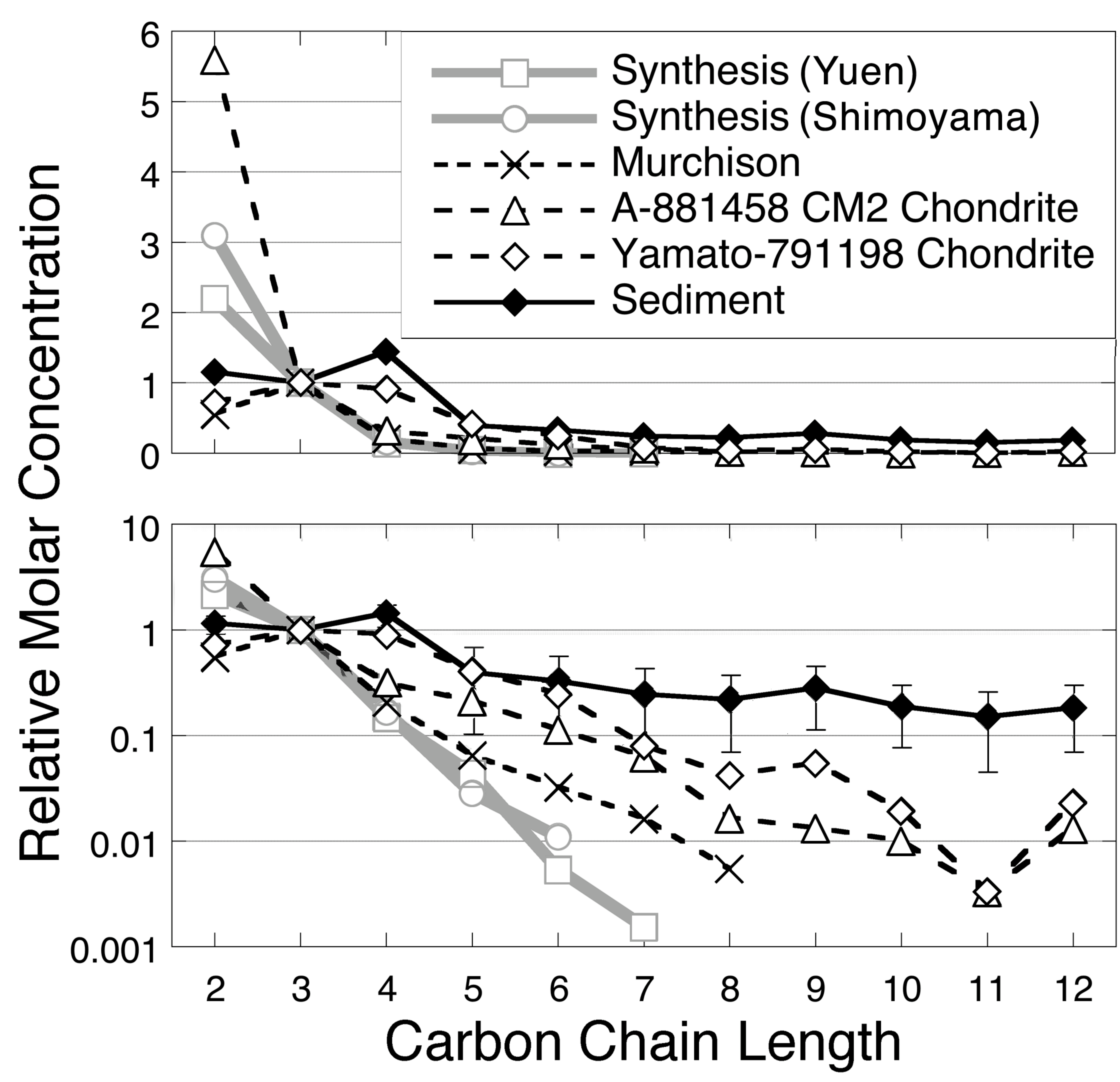}
\caption{Relative concentrations of low-carbon number monocarboxylic
acids from five abiotic sources compared to the average of 12
measurements in sediments.  The data are plotted again on a semilog
scale below to show the trend of decreasing concentration after C6:0 in
abiotic samples.  Synthesis curves represent two individual experiments (Shimoyama et al. 1994; Yuen et al. 1991).}
\label{Figure 2}
\end{figure}

\begin{figure}
\centering
\includegraphics[width=10cm]{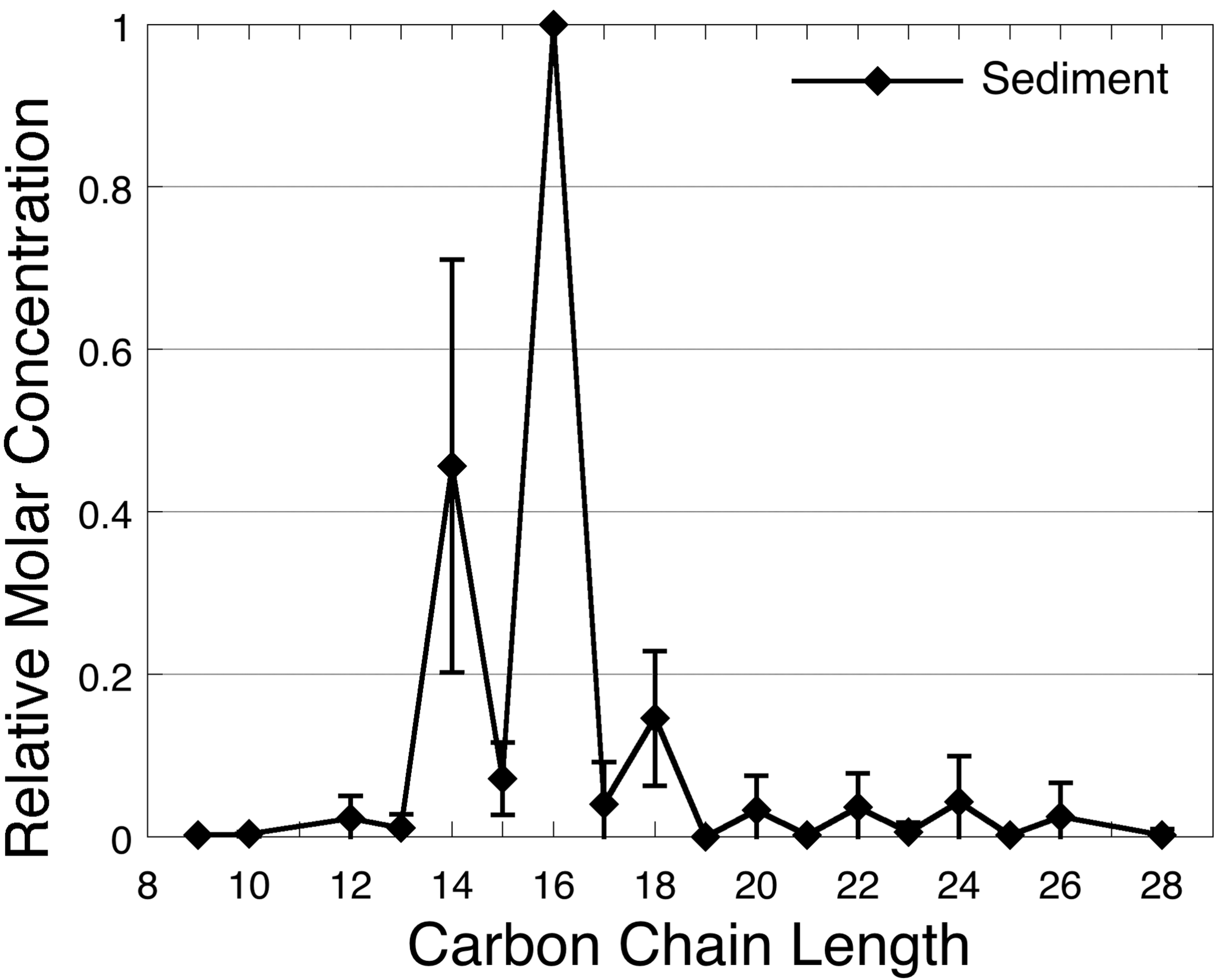}
\caption{Curve representing the average pattern of long-chain
monocarboxylic acids found in terrestrial sediments ($n=26$), showing an apparent 
bias towards even-numbered carbon chains not seen in abiotic syntheses.}
\label{Figure 3}
\end{figure}

In terrestrial samples, however, the higher molecular weight acids C14:0
through C32:0 are present in abundance. As shown in Figure 3,
environmental samples exhibit a consistent pattern that includes a primary
peak at C16:0 with smaller peaks favoring even-carbon-number acids. This
pattern cannot be explained by formation thermodynamics, but rather
reflects a pair of biological constraints. A fitness criterion is
present: carbon chain lengths near 16, in both single- and double-chain
amphiphiles, are ideal for the formation of stable, non-porous, and
flexible bilayers in aqueous solution at biologically relevant ranges of
temperature and pH \cite{HargreavesWR_DeamerDW:1978,DeamerDW:1999,GebickiJM_HicksM:1976}, 
and are therefore critical for membrane structure in
a water-based biosphere. The predominance of even-numbered carbon chains
reflects processes specific to terrestrial biochemistry: the primary
fatty acid biosynthesis cycle terminates at C16:0 (palmitate), from
which other fatty acids are synthesized through lengthening or
shortening cycles that operate in two-carbon steps. A bias toward
even-numbered carbon chains is considered diagnostic of biosynthesis and
has been used to identify terrestrial contamination in meteorite sources
\cite{NagyB_BitzSMC:1963}. The slight increase in concentration of C12:0 seen
in two of the meteorites in Fig. 2 probably represents low-level
terrestrial contamination.

Because all carboxylic acids from C2 through C28 are found in measurable
concentrations in both biotic context and in abiotic context (when FTT syntheses
are included), the Lego Principle \cite{McKay:2004} - that biological samples include only a subset of the 
possible molecules in a family -  is not observed in monocarboxylic
acids.   However, because the shapes of the abundance curves are so different in the
two contexts, an MADB is still observable in the sense originally described by 
Lovelock \cite{Lovelock:1965}.   

\subsection{Digital Life}

\subsubsection{Monomer Abundances in Terminal Populations}

In Avida, two processes are present that we may consider ``abiotic.'' 
First, we initalize the population with a hand-written, progenitor
genotype.  Second, mutations in Avida result in the replacement of an
instruction with a new one, randomly chosen from the library with equal
probability for each possible instruction.  If selection did not affect
the distribution of instructions, we would expect one or both of these
effects to constrain the evolved abundance.  If the initial conditions
imposed by the ancestor constrained the evolved population, we would
expect the final abundance to reflect the ancestor's distribution.  If
mutation were the dominant effect, as we would expect over long periods
of time, the final population would show each of the instructions in
equal proportion, demonstrating that the genotypes had incorporated the
instructions at the same rate with that they appreared in mutation.
Thus, the abiotic expectation in Avida, analogous to the Synthesis
curves of Figs. 1 and 2, is a flat line.

We observe in nearly all Avida runs, regardless of initial conditions or
variables, a specific and consistent pattern of evolved instruction
abundance that reflects neither of the abiotic constraints as a dominant
feature. Figures 4a and 4b compare bulk instruction frequency curves to
the abiotic expectation for populations from the two progenitor
organisms in Table 1. Experiments were conducted at a number of
different mutation rates, where $\mu$ represents the probability that an instruction is mutated
when it is being copied into a progenitor's genome (the "per-site copy
mutation" rate).  While some variability is
present as conditions change, many of the general features of the
pattern are conserved and all evolved patterns are distinguishable from
the abiotic baseline. The relative abundances observed in our
experiments differ not only in their mean, but also in their variance.
Instructions critical to an organism's survival (such as COPY and
DIVIDE) are usually present at predictable levels (here, one per
genome), and therefore show little variability. Optional instructions
such as NAND, on the other hand, are expressed at high but variable
concentrations as organisms in different competitive environments use
them to complete different mathematical tasks (see ``The Avida
Environment'' in Materials and Methods). Instructions that are frequently
lethal when they appear as mutations (e.g., RETURN and JUMP-F) are
suppressed, and the useless but non-fatal instruction NOP-X appears at a
low but persistent level.

\renewcommand{\baselinestretch}{1.0}
\begin{table}
\centering
\begin{tabular}{l||l|l||l|l}
\multicolumn{1}{c||}{\textbf{Progenitor}}& \multicolumn{2}{c||}{\textbf{Alternate Progenitor}}
& \multicolumn{2}{c}{\textbf{Typical Evolved Avidian}}\\
\hline
1 SEARCH-F  & 1 CALL   & 29 DEC & 1 SEARCH-F & 29 NOP-C\\ 
2 NOP-A & 2 NOP-A & 30 NOP-C & 2 NOP-A & 30 PUT\\ 
3 NOP-A & 3 NOP-A & 37255 & 3 NOP-A & 31 NOP-X\\ 
4 ADD  & 4 SHIFT-L  & 32 NOP-C & 4 ADD & 32 DIVIDE\\ 
5 INC  & 5 ALLOCATE  & 33 PUSH & 5 ADD & 33 PUT\\ 
6 ALLOCATE  & 6 PUSH    & 34 NOP-C & 6 GET & 34 SHIFT-R\\ 
7 PUSH  & 7 SWAP-STK & 35 RETURN & 7 ALLOCATE & 35 DEC\\ 
8 NOP-B & 8 PUSH    & 36 NOP-X & 8 NAND & 36 PUT\\ 
9 POP & 9 NOP-B   & 37 NOP-X & 9 PUT & 37 NOP-B\\ 
10 NOP-C & 10 POP    & 38 NOP-X & 10 NOP-C & 38 COPY\\ 
11 SUB & 11 NOP-C   & 39 NOP-X & 11 PUT & 39 INC\\ 
12 NOP-B  & 12 POP     & 40 NOP-X & 12 NAND & 40 IF-N-EQU\\ 
13 COPY & 13 NOP-B   & 41 NOP-X & 13 GET & 41 NOP-A\\ 
14 INC & 14 IF-N-EQU & 42 NOP-X & 14 NAND & 42 JUMP-B\\ 
15 IF-N-EQU & 15 NOP-A & 43 NOP-X & 15 GET & 43 NOP-A\\ 
16 JUMP-B & 16 CALL      & 44 NOP-X & 16 PUSH & 44 GET\\ 
17 NOP-A & 17 NOP-A & 45 NOP-X & 17 PUT & 45 NOP-B\\ 
18 DIVIDE  & 18 NOP-B & 46 NOP-X & 18 POP & 46 NOP-B\\ 
19 NOP-B & 19 DIVIDE & 47 NOP-B & 19 NOP-X & 47 JUMP-B\\ 
20 NOP-B & 20 JUMP-F & 48 NOP-B & 20 IF-N-EQU & \\ 
 & 21 NOP-B  & 49 SEARCH-B   & 21 NAND & \\ 
 & 22 NOP-C & 50 NOP-A   & 22 IF-BIT-1 & \\ 
 & 23 COPY & 51 NOP-B & 23 NAND & \\ 
 & 24 INC & 52 RETURN & 24 GET & \\ 
 & 25 POP   & 53 NOP-X & 25 NAND & \\ 
 & 26 NOP-C & 54 NOP-C & 26 NAND & \\ 
 & 27 DEC & 55 NOP-A & 27 NOP-C & \\ 
 & 28 NOP-C &  & 28 PUT & 
\end{tabular} 
\label{Table 1}
\caption{Examples of three Avida organisms. The progenitors, used to
seed initial populations, were hand-coded and are poorly adapted to the
Avida environment.  The ``evolved avidian'' is an example genome that
arose approximately 1000 generations into a typical run seeded with the
standard progenitor.}
\end{table}
\renewcommand{\baselinestretch}{1.5}

\begin{figure}
\centering
\includegraphics[width=14cm]{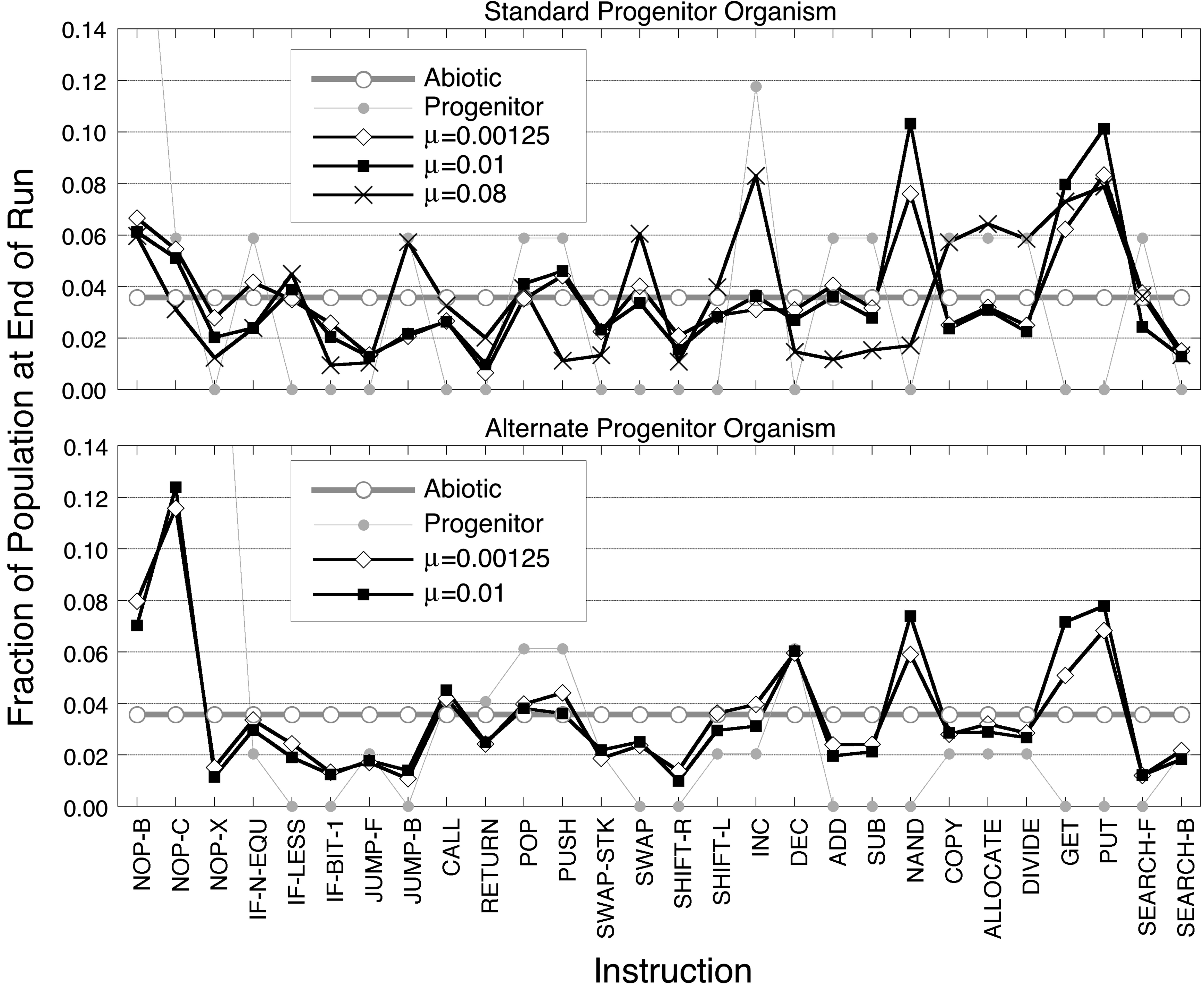}
\caption{Relative abundance curves for several evolved
populations.  Note how a wide range of mutation rates ($\mu$=0.00125 to
0.01) produces nearly identical distributions.  Tests at numerous
intermediate values of $\mu$ showed similar results (not shown).  
Near-lethal mutation rates ($\mu$=0.08) and alternate progenitor organisms
produce distinct distributions that share some features with the
``standard'' distribution.  $n=25$ for each curve.}
\label{Figures 4a and 4b}
\end{figure}

The distribution of relative abundances is essentially consistent over a
large range of mutation rates (factor of 32), and changes only as the
mutation rate becomes so high that the majority of offspring contain at
least one mutation. At a near-lethal mutation rate of $\mu$=0.08, a change
in survival strategy generates a dramatic shift in the biosignature.
Organisms abandon mathematical tasks and opt for very short genomes in
order to maximize the likelihood of spawning genetically pure offspring.
Mathematics-related instructions (NAND) are suppressed, while
instructions that generally only appear once per genome (COPY, ALLOCATE,
DIVIDE) are present at a higher concentration in the population as a
whole because of the shorter average genome length. Other features, such
as suppression of often-lethal mutations (JUMP-F, RETURN) are retained.

The observed pattern of evolved instruction abundance retains most of
its salient features when we seed the population with a drastically
different progenitor organism (See Table 1), demonstrating that the
constraints of adaptation generally overwhelm the effect of initial
conditions. As runs progress, the signatures of populations descending
from the two disparate progenitors tend to converge, as some features are
conserved among well-adapted populations.   Most populations tend to express
high frequencies of NAND, since it is essential for solving mathematical tasks
and therefore high fitness.  

There are some features of the abundance profile that Avidians tend to retain 
from their ancestor.  For example, descendants of the standard progenitor 
tend to use JUMP-B (jump backward) for loop construction, while descendants of the alternate
progenitor tend to use CALL/RETURN (call subroutine and return from subroutine) for the 
same basic function.   Likewise one clade tends to use decrementing loop counters
with DEC while the other uses INC.  These local fitness maxima represent distinct genotypes that are fairly well adapted themselves but cannot be  derived from each other via evolution because the intermediate mutations
required to convert one genotype into the other would be invariably
fatal. For example, the two progenitors differ in implementing their
essential ``copy loop'': one uses the INC (increment) operator while the
other uses DEC (decrement).  Changing INC to DEC in the copy loop would
require several simultaneous mutations, any one of which alone would be
fatal. As a result, this transformation is never seen in practice.
Though INC and DEC may serve other functions in an organism, evolved
populations tend to express the operator used by their ancestor's copy
loop. This effect is evident in Fig. 4.

\begin{figure}
\centering
\includegraphics[width=15cm]{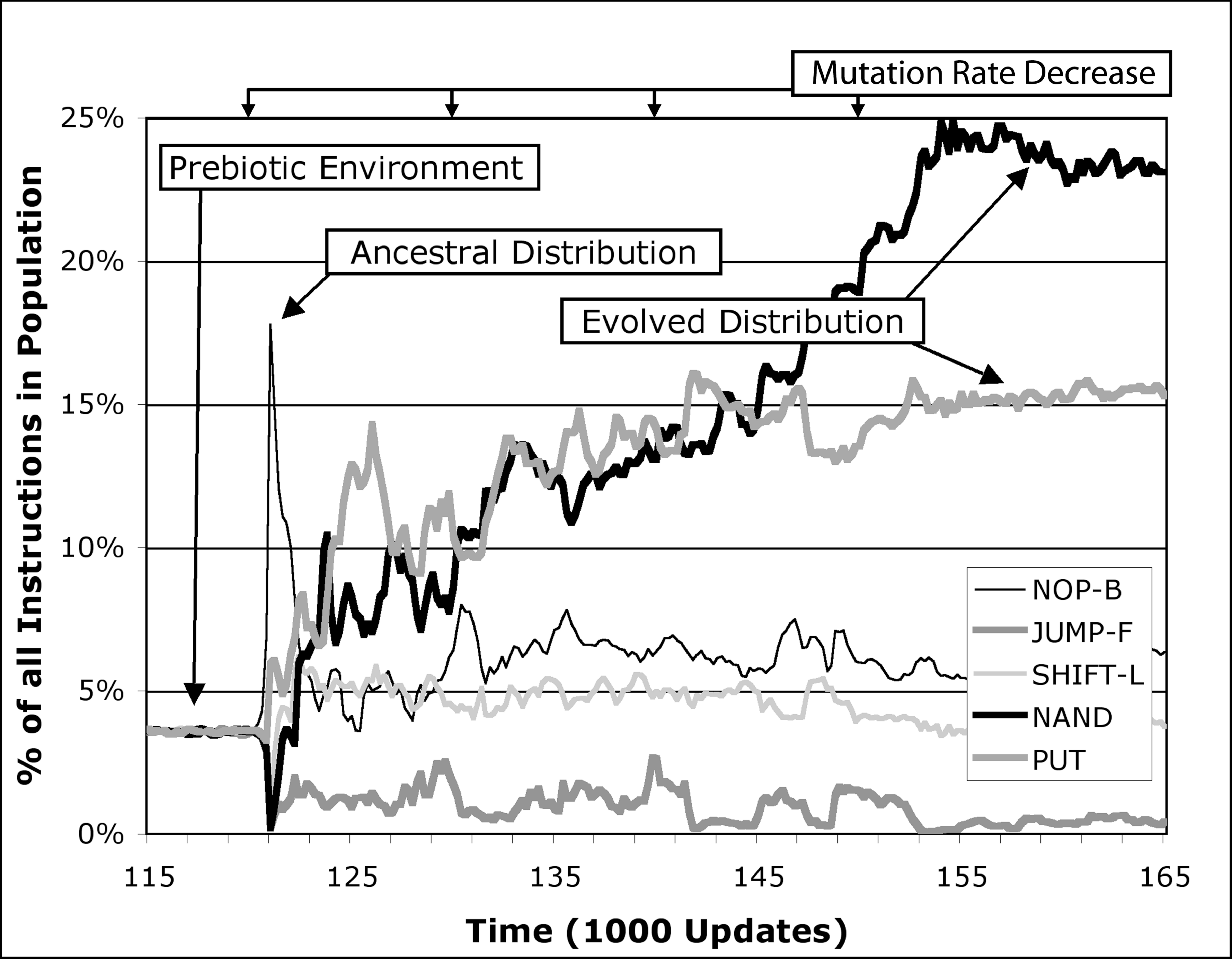}
\caption{Abundance vs. time for six instructions as an incident lifeform
populates a previously abiotic environment.  The ancestor's most
abundant instruction is NOP-B, while the adapted population's most
abundant instruction is NAND.  All instructions were present in equal
proportion in the initial, mutation-dominated environment that did not
contain any replicating organisms.  A single standard Avida ancestor was seeded
into the population every 200 updates.   Point mutation (``cosmic ray'') rate was stepped down
progressively every 10,000 updates from the start of the experiment until update
150,000.  At 120,000 updates the mutation rate dropped from the lethal level of
2.67\e{-3} mutations per update per site to a nonlethal level of 2.00\e{-3} mutations
per update per site.  After update 120,000, one of the ancestors
was able to replicate and establish a population before being destroyed by
point mutations.}
\label{Figure 5}
\end{figure}

\subsubsection{Time-Evolution of a Biosignature}
 
When viable self-replicators begin to dominate a previously lifeless
Avida grid, their signature frequency of instructions quickly overwhelms
the pre-existing random distribution. Figure 5 shows the time-varying
abundances of several instructions in one such run where sequences capable of self-replication were consistently re-injected into the population at a small rate, while sequences were bombarded with a high rate of point mutations per-site that is independent of the copy-process of replication, that is, the mutations were of the ``cosmic ray" type.
At the outset, all
instructions were present in equal abundance as the cells contained
randomized, nonviable code because the injected replicators could not take hold. This gives rise to exactly the hypothetical abundance pattern
used as an abiotic baseline in Fig. 4. When the point mutation rate was
decreased to a non-lethal level, replication immediately became the dominant
process. As the population of replicators filled the available space, the distribution of
instructions in the ancestor's genome began to dominate, but was quickly
replaced as higher-fitness organisms evolved. A recognizable
distribution evolves with the same features we see in Fig. 4, such as
high NAND concentration but suppressed JUMP-F.  The distribution
stabilizes after a few hundred generations, even though the organisms
themselves were still evolving.

We tracked all 28 instructions and allowed the population to evolve for
several thousand generations beyond the point shown in Fig. 5.
Eventually, we increased the mutation rate to its previous level, and
the population died.  At that point, the bombardment of point mutations
rapidly returned the environment to its prebiotic state. Digital movie
files showing the time evolution of all 28 instructions for the full
time course of the experiment shown in Fig. 5., and the development of
one run used to generate Fig. 4, may be found in the online supplemental
information.

\section{Discussion}

We believe the monomer abundance distribution biosignature is a universal feature of
evolving systems: rapid synthesis of a select subset of compounds by a
living metabolism should inevitably leave a biosphere in notable
disequilibrium with respect to a sterile planet of the same
geochemical composition.   If we consider any particular monomer family,
it is unlikely that thermodynamics would constrain abiotic synthesis to
precisely the same rates as biotic synthesis.  Indeed, if abiotic
synthesis did produce monomers in the same concentrations as evolved
biosynthesis, then biosynthesis would not be necessary, selection could
not improve on abiotic geochemistry, and we question whether such a
circumstance could in fact be described as ``life.''

The ``lock-in'' effect where some features of the ancestral avidian 
survive despite the availability of equivalent instructions is analogous to 
the persistent use by terrestrial biology of a subset of the amino acids,  selected 
early in life's evolution, that have complementary properties.  For example, it 
is essential that we have amino acids with both hydrophilic and hydrophobic side-chains,  
but it may not have been necessary that life uses exactly the sets we have.  The possibility of
early life having  being initiated with a different set of amino acids that nonetheless
span the necessary and complementary functions is analogous to the early 
lock-in of CALL/RETURN rather than JUMP-B for looping.  

Furthermore, the MADB should be detectable whenever life is present.  While 
both abiotic and biotic syntheses may be present in an environment, we may 
reasonably expect biosynthesis to
dominate most molecular synthesis as it does in the terrestrial
biosphere. The coupling of energy from metabolism to catabolism, the
manufacture of catalysts (enzymes), and the ability to concentrate
reactants inside cells conspire to generate vast increases in synthesis
rates relative to abiotic chemistry.  As a result, even small quantities
of extant life should leave a detectable chemical signature.

Understanding evolution's effects on the chemical environment resulting
from selection operating on metabolizing organisms carries
significance both for our general understanding of life and for the
search for extraterrestrial life.  It is essential for the latter
because we cannot predict the specific biochemistry of the putative life
for which we are searching.  A hypothetical extraterrestrial biochemistry
might use a different set of amino acids, or use them in different
abundances.  But while such a pattern may not look like the terrestrial
data in Fig. 1, it would be unlikely to match the consistent
pattern seen in abiotic syntheses. Likewise, extraterrestrial fatty acid
biosynthesis could conceivably use cycles that operated in one- or
three-carbon increments, resulting in a different distribution of peaks.
If the organisms used ammonia, methane, or some other primary solvent
rather than water, the adaptive peak for stable membranes might not be
C:16. In either case, the unique signature of this exotic biochemistry
would reflect the adaptive constraints of these organisms' biosphere and
could be very unlike the ``Synthesis'' curves in Figure 2.

It is of course also possible that an extraterrestrial biosphere might not 
employ amino- and carboxylic acids at all.   Although some argue
that the same families of molecules seen in terrestrial biology (proteins, fats,
sugars, nucleic acids, etc.) are likely to be found in carbon-based life everywhere 
\cite{PaceNR:2001}, an unbiased search should include the possibility
of life using alternate chemistry.   Doing so would mean generating thorough
abiotic profiles of a wide variety of chemical families, and equipping 
a probe or measurement device with the ability to assay numerous compounds
in each family to detect abundance profiles deviating from the range of 
abiotic measurements.

A contained digital system like Avida has several advantages for
studying the MADB hypothesis. The course of
evolution can be studied repeatedly to provide statistical significance
to dynamics that usually only yield singular, history-dependent event.
With terrestrial life, we can only examine one example of an evolved
biochemistry: we cannot ``start life over'' on Earth and measure the
concentrations of amino or carboxylic acids in each resulting biosphere,
and therefore cannot prove that the particular concentration patterns we
observe are the result of evolutionary selection's effect on biochemistry rather than
happenstance. In Avida, we can repeat the process indefinitely to obtain
statistics about how organisms adapt. In addition, we have complete
control over initial conditions, allowing us to accurately characterize
their impact on the result. Finally, Avida implements the replication and 
evolution aspects of life in a system otherwise completely unrelated to
terrestrial biochemistry, giving us an opportunity to study a
biosignature protocol in a Earth-independent setting.

\section{Conclusions}

The relative distribution of abundances of a set of monomers is a strong
biosignature that may show promise for detecting alien
biochemistries where sufficient population exists to imprint a chemical 
signature. The constraints of adaptation, combined with the growth of biomass,
overwhelm the abiotic chemistry present in the environment as organisms 
expend energy to build complex, low-entropy structures.  The similarity between
laboratory synthesis and meteorite samples is remarkable since
meteorite samples may have formed under conditions very different from
laboratory syntheses and have aged under planetary and/or deep space
conditions for thousands or millions of years before collection and
analysis.

Using the digital life platform Avida, we have demonstrated that the monomer
abundance biosignature phenomenon may be germane to any living
system, and have studied the robustness of the marker across a range of
mutation rates and seed organisms. While an exhaustive analysis of
initial conditions cannot be performed for this simple model life-form, 
the robustness of discrimination we see given the variety of
evolutionary paths and organisms produced in these experiments bodes
well for the universality and applicability of the MADB.

It may be possible to look for instances of the monomer abundance
biosignature simply by empowering a life-detection system with a few
basic models (derived theoretically and/or empirically) of expected
abiotic chemical abundances, and then examining samples for patterns
that differ significantly. Strong deviations from thermodynamic/kinetic
expectations, i.e., the appearance of unusual concentrations of any high
formation energy, low probability compounds, could signal that a site or
sample may contain life and is worthy of further investigation.  This
strategy poses significant challenges, but ultimately frees the search
for extraterrestrial life from biases centered in our present
understanding of terrestrial biochemisty.

Creating accurate models of expected conditions will be difficult, as
unfamiliar local effects and boundary conditions (e.g., excessive
concentrations of a certain mineral, or unusual temperatures) could
change the chemical species present in a sample and cause a false
positive. However, we may find some safe generalizations (for example,
that lysine, arginine, and histidine will always be absent or low in the
absence of life) that enable astrobiologists to build reliable
experiments.

\section{Acknowledgements}

The research described in this work was carried out in part at the Jet
Propulsion Laboratory, California Institute of Technology, under a
contract with the National Aeronautics and Space Administration (NASA),
with support from the Director's Research and Development Fund (DRDF)
and from the National Science Foundation under contract No. DEB-9981397.
We thank Claus Wilke, Ronald V. Dorn III, and Diana Sherman for
discussions. Finally, we are grateful to three anonymous reviewers for extensive and constructive comments on the manuscript. 

 \bibliography{biomarkers,alife}
 \end{document}